\documentclass[12pt,a4paper]{article}

\RequirePackage{amsmath,amssymb,graphicx,txfonts,mathrsfs}
\usepackage[section] {placeins}		
\usepackage[]{natbib}

\bibpunct{(}{)}{;}{a}{,}{;}

\usepackage{lscape}

\newcommand{\E}{{\operatorname{E}}}

\def\K{\scriptscriptstyle K}
\def\H{\scriptscriptstyle H}

\usepackage{caption}
\begin{document}

\baselineskip 8mm

\title{
On predictability of ultra short AR(1) sequences
\footnote{We acknowledge provision of ICTsupport and computing resources by Curtin IT Services http://cits.curtin.edu.au. Curtin Information Technology Services (CITS) provides Information and Communication Technology systems and services in support of Curtin's teaching, learning, research and administrative activities. We acknowledge use of computing resources from the NeCTAR Research Cloud http://www.nectar.org.au. NeCTAR is an Australian Government project conducted as part of the Super Science initiative and financed by the Education Investment Fund.
This work was
supported by ARC grant of Australia DP120100928}}
\author{\sc
Nikolai Dokuchaev \footnote{Corresponding author. Department of Mathematics and Statistics,
Curtin University,  GPO Box U1987, Perth, 6845 Western Australia;
email: N.Dokuchaev@curtin.edu.au} \
and \   Lin-Yee Hin
\\
 { Department of Mathematics \& Statistics, Curtin
University,}\\ {  GPO Box U1987, Perth, 6845 Western Australia}}

\maketitle

\thispagestyle{empty}

\begin{abstract}
This paper addresses short term forecast of ultra short AR(1) sequences (4 to 6 terms only) with a single structural break at an  unknown time and of unknown sign and magnitude.
As prediction of autoregressive processes requires  estimated coefficients, the efficiency of which relies on the large sample properties of the estimator,
it is a common perception that
prediction is practically impossible
for such short series with structural break.
However,
we obtain a heuristic result that
some universal predictors represented
in the frequency domain
allow certain predictability
based on these ultra short sequences.
The predictors that we use are universal in a sense that they are not oriented on particular types of autoregressions and do not require explicit modelling of structural break.
The shorter the sequence,
the better
the one-step-ahead forecast performance of
the smoothed predicting kernel.
If the structural break entails a model parameter switch from negative to positive value,
the forecast performance of
the smoothed predicting kernel
is better than that of
the linear predictor that utilize AR(1) coefficient estimated from the ultra short sequence without taking the structural break into account
regardless whether
the innovation terms in the learning sequences are constructed from
independent and identically distributed random  Gaussian or Gamma variables,
scaled pseudo-uniform variables,
or
first-order auto-correlated Gaussian process.
\end{abstract}


\newpage

 \setcounter{page}{1}

\section{Introduction}

In this paper, we readdress the problem of one-step-ahead  forecast of a first order autoregressive process, AR(1), with one structural break, i.e., a permanent change, in the AR(1) model parameter.
Specifically,
we consider the scenario where the learning sequence, i.e., the segment of time series process used for model estimation and forecast, is very short,
and
where the structural break occurs at a random time point in the learning sequence. \par

Forecasting autoregressive process is a well developed area with well known results.
One strand of literature addresses this problem via
the time series models
that are primarily specified from the time-domain modelling perspective
\citep[see, among many others,][and the references therein]{Box:Jenkins:1976,
Abraham:Ledolter:1986,
Stine:1987,
Cryer:Nankervis:Savin:1990,
Cotez:Rocha:Neves:2004,
Hamilton:1994,Xia:Zheng:2015},
or
via
the exponential smoothing
and
the filtering techniques
constructed
based on state-space approach
where
the smoothers and filters are
primarily characterized in
the time-domain
\citep[see, e.g.,][and references therein]{Roberts:1982,
Williams:1987,
Paige:Saunders:1977,
Chatfield:Yar:1988,
Ord:Koehler:Snyder:1997,
Chatfield:Koehler:Ord:Snyder:2001,
Hyndman:Koehler:Snyder:Grose:2002,
Bermudez:Segura:Vercher:2006,
Hyndman:Koehler:Snyder:Ralph:2008}.
A separate yet related strand of literature
addresses this problem via
smoothing and filtering techniques
where
the smoothers and filters are
primarily characterized in
the frequency-domain
\citep[see, e.g.,][and references therein]{Cambanis:Soltani:1984,
Ledolter:Kahl:1984,
Lyman:Edmonson:2001,
Dokuchaev:2012,
Dokuchaev:2014,
Dokuchaev:SP:2016}.
In this paper, we address this problem via
the convolution of
a near-ideal causal smoothing filter
and
a predicting kernel
that are primarily characterized in
the frequency-domain
\citep{Dokuchaev:2012,
Dokuchaev:2014,
Dokuchaev:SP:2016}.   \par

Many strategies have been proposed
to address the
practical concern
of possible model parameters structural break in the learning sequence
that may compromise modelling efficiency and forecast performance of the time series model
\citep[see, among many others,][and the references therein]{Bagshow:Johnson:1977,
Sastri:1986,
Andrews:1993,
Ba1:Perron:1998,
Ba1:Perron:2003,
Pesaran:Timmermann:2004,
Clements:Hendry:2006,
Davis:Lee:RodriguezYam:2006,
Lin:Wei:2006,
Kim:Koh:Boyd:Gorinevsky:2009,
Rossi:2013,
Pesaran:Pick:Pranovich:2013}.
Implementation of these strategies
require the availability of
learning sequences
that are considerably longer that those considered in this paper.
We cite a few examples.
The method proposed by \cite{Ba1:Perron:1998,Ba1:Perron:2003} to estimate the timing of structural break requires at least 10 observations on either side of the break.
\cite{Pesaran:Timmermann:2007}
simulated random processes that each contain 100 to 200 observations
to mimic learning sequences with structural break
in AR(1) model parameter
in order to assess the performance of
their proposed
set of cross-validation and forecast combination procedures
that use pre-break and post-break data to perform time series forecast.
\cite{Giraitis:Kapetanios:Price:2013}
simulated time series processes that each contain 200 observations
to mimic learning sequences with structural break
in the mean of the simulated random processes
in order to assess the performance of
their proposed
one-step-ahead forecast algorithms based on adaptive linear filtering.
  \par

In this paper,
we consider the scenario when the learning sequence only contain 4 to 6 data points,
and, as such, are
too short to effectively apply structural break timing estimation strategies,
and to efficiently estimate pre-break and post-break AR(1) model parameters.
We consider a family of linear filters
proposed by \cite{Dokuchaev:SP:2016}
where
the impulse response function is
obtained by inverse Z-transform
of
the product between
the transfer function of a family of near-ideal causal smoothers \citep{Dokuchaev:SP:2016}
and
the transfer function of a family of predicting kernels
\citep{Dokuchaev:2014}.
The Monte Carlo experiments reported in
\cite{Dokuchaev:SP:2016}
have demonstrated clear advantage of
using
the
convolution of the near-ideal causal smoother and the predicting kernel
compared to
using
the
predicting kernel alone
to generate the impulse response function
for the linear predictor
in terms of one-step-ahead forecast performance of AR(2) processes without structural break.
However,
their relative forecast performance
have never been assessed in the context of
AR(1) processes with a single, unknown, random time point structural break in a very short learning sequence.
Additionally,
numerical experiments reported in
\cite{Dokuchaev:SP:2016}
utilize 100 observations in the learning sequence.
This begs the question whether the proposed
linear predictor will perform well
in the context of a very short learning sequence with structural break.
This paper seeks to close this gap in the literature. \par

Following the choice of benchmark used in,
among others,
\cite{Pesaran:Timmermann:2007}
and
\cite{Giraitis:Kapetanios:Price:2013},
we use the one-step-ahead forecasts from an AR(1) model that ignores structural break and
utilize all observations, pre-break and post-break, as our  benchmark
since this is an appropriate model to use in situations with no breaks.
The main contribution of this paper
is
our demonstration via simulation experiment that
the one-step-ahead forecast performance of
this family of linear filters
is
better than
that of
our chosen benchmark.
Additionally,
its performance is comparable to
that of the one-step-ahead forecasts
from
an AR(1) model with the same model parameter as the synthetic AR(1) model parameter used to simulate the post-break random process. \par

The rest of the paper is as follows.
Section \ref{Problem:setting}
details the problem formulation.
Section \ref{Simulation:experiment}
presents the Monte Carlo simulation results.
Section \ref{Conclusions} concludes. \par

\def\T{\theta}
\section{Problem setting}
\label{Problem:setting}

Consider a stochastic discrete time process described by AR(1) autoregression
\begin{align}
x(t)
&=
\beta(t) \; x(t-1)
+ \sigma \; \eta(t) \;,
\quad t = 0, \ldots, d-1 \;\quad x(-1)=0,
\label{x(t):sequence} \\
\beta(t)
&=
\begin{cases}
\beta_1 \;, &t < \T \;,\\
\beta_2 \;, &t \ge  \T \;.
\end{cases}
\nonumber
\end{align}
where
$\beta_1 \in
(\beta_{\scriptscriptstyle min},\beta_{\scriptscriptstyle max})$,
$\beta_2 \in
(\beta_{\scriptscriptstyle min},\beta_{\scriptscriptstyle max})$,
$\beta_{\scriptscriptstyle min} < \beta_{\scriptscriptstyle max}$,
$| \beta_{\scriptscriptstyle min} | < 1$,
$| \beta_{\scriptscriptstyle max} | < 1$,
$\sigma \in (0,\infty)$,
and
$\eta(t)$ is the innovation term of the time series.
This model features a single random structural break to take place at a random time $\T$ with the values in the set $\{ 1,\ldots,d-2 \}$.
We assume that $\eta(t)$ are mutually independent for all $t$ and  independent on $\eta$.

We consider predicting problem for this process in the case where an ultra short sequences with no more than
six data points are available.
\subsection{Special predictors }
We investigate the performance of linear time-invariant predictors  with an output
\begin{equation}
y(t) = \sum_{\tau =0}^{t}
h(t-\tau) \; x(\tau), \quad  t\le d-1,
\label{y(t):sequence}
\end{equation}
where $d\le 6$.
The process $y(t)$ is supposed to  approximate the process $x(t+1)$,
i.e., \eqref{y(t):sequence} represents a one-step-ahead predictor.
The predictor is defined by a impulse response function
$h: \mathbb{Z} \rightarrow \mathbb{R}$, where
$\mathbb{Z}$ is the set of integers, and
$\mathbb{R}$ is the set of real numbers.

In our experiments, we calculate predicting kernels $h$  via   their
Z-transforms that are represented explicitly, such that
\begin{equation}
h(t)
=
\frac{1}{2\pi}
\int_{-\pi}^{\pi}
H\left(e^{i\omega}\right) e^{i \; \omega \; t} d\omega \;,
\quad t \in \mathbb{Z}.
\end{equation}
Here  complex-valued functions
$H:\mathbb{C} \to \mathbb{C}$ are transfer functions of the corresponding predictors.

 In our experiments, we used two different transfer functions
\begin{equation}
H(z) = K(z) \;,
\label{prediction:kernel:predictor}
\end{equation}
and
\begin{equation}
H(z) = K(z) \; F(z).
\label{smooth:and:prediction:kernel:predictor}
\end{equation}  Here $z\in\mathbb{C}$,
\begin{equation}
K(z)
=
z \left( \;
1 -
\exp\left[
- \frac{\gamma}{z+1-\gamma^{-r}}
\right] \;
\right).
\label{predictor:transfer:function}
\end{equation}
The function $K(z)$ is the transfer function of an one-step predictor  from  \cite{Dokuchaev:SP:2016};  $r > 0$, $\gamma > 0$ are the parameters.

In (\ref{smooth:and:prediction:kernel:predictor}),
\begin{align}
F(z)
&= \left(
\exp
\frac{(1-a)^p}{z+a}
+
G(z)
\right)^m \;, \label{smooth:transfer:function} \\
G(z)
&=
-\xi(a,p)
+ \frac{\gamma(a,p)}{N}
\left( (-1)^Nz^{-N} - 1 \right) \;, \nonumber \\
\xi(a,p)
&=
\exp[-(1-a)^{p-1}] \;, \nonumber\\
\gamma(a,p)
&=
| 1 - a |^{p-2} \xi(a,p) \;. \nonumber
\end{align}
Here  $a\in (0,1)$, $p\in (1/2,1)$, $m\ge 1$, and $N\ge 1$, are the parameters, $m,n\in \mathbb{Z}$.
The function $F(z)$ is the transfer function for a smoothing filter introduced in  \cite{Dokuchaev:SP:2016}.

It can be noted that these linear predictors were constructed for semi-infinite one-sided sequences,
since the corresponding kernels $h(t)$  have infinite support on $\mathbb{Z}$. In theory, the performance of these predictors is robust with
respect to truncation; see the discussion on robustness in  \cite{Dokuchaev:2012}
 and \cite{Dokuchaev:SP:2016}.   However, we found, as a heuristic result of this paper,
that their application to the ultra short series also brings some positive result, meaning
that these sequences feature  some  predictability.
Worthy of note again is that implementation of these predictors does not involve explicit
modelling of and  adjustment for structural break, including break time and magnitude,
Moreover, these  predictors  do not even require that the underlying process is an autoregressive process or any other particular kind of a processes. \par

\subsection{Comparison with other predictors}
We compare the performance of our predictors with
an ``ideal'' linear predictor
\begin{equation}
y_{\scriptscriptstyle ideal}(d-1)
=
\beta(d) \; x(d-1) \;,
\label{optimal:predictor}
\end{equation}
where $\beta(d)=\beta_2$ is the post-break AR(1) model parameters that generate the post-break
observations $x(d-1)$ and $x(d)$.
This predictor  is not feasible unless $\beta(d)$ is supposed to be known.
In our setting,
$\beta(d)$ is unknown and has to be estimated from the observations.  We will use the performance of this predictor as a benchmark.\par

Additionally. we will compare the performance of our predictors with  the performance of the predictor
\begin{equation}
y_{\scriptscriptstyle AR(1)}(d-1)
=
\hat{\beta}(d) \; x(d-1)  \;,
\label{AR1:predictor}
\end{equation}
where
$\hat{\beta}(d)$
is estimated
by fitting an AR(1) model to the sequence
$\left\{ x(\tau) \right\}_{\tau=0}^{d-1}$ that involve pre-break and post-break observations
using the build-in function {\tt ar.ols()} in the {\tt R} computing environment \citep{R:2016}
implementing the
ordinary least squares model parameter estimation strategy
\cite[pp. 368-370,][]{Luetkepohl:1991}.
This is an appropriate model estimation procedure to use if the sequence
$\left\{ x(\tau) \right\}_{\tau=0}^{d-1}$
does not contain structural break.
By choosing this predictor as the benchmark for our numerical experiment,
we seek to address
the question of
how costly is it to ignore breaks when
performing one-step-ahead forecasting the direction of a time series
using
the prediction algorithms considered,
i.e.,
\eqref{prediction:kernel:predictor},
\eqref{smooth:and:prediction:kernel:predictor},
and
\eqref{optimal:predictor},
relative to
\eqref{AR1:predictor}.

\section{Simulation experiment}
\label{Simulation:experiment}

We perform simulation experiments
to investigate
the one-step ahead forecast performance of
\eqref{prediction:kernel:predictor},
\eqref{smooth:and:prediction:kernel:predictor},
and
\eqref{optimal:predictor},
relative to
\eqref{AR1:predictor}
in predicting $x(d)$,
given $\left\{ x(\tau) \right\}_{\tau=0}^{d-1}$
simulated from \eqref{x(t):sequence}
using
four different specifications of $(\beta_1,\beta_2)$
\begin{enumerate}
\item $\beta_1 \in (0,1), \beta_2 \in (0,1)$,

\item $\beta_1 \in (-1,1), \beta_2 \in (-1,1)$,

\item $\beta_1 \in (-1,0), \beta_2 \in (0,1)$,

\item $\beta_1 \in (0,1), \beta_2 \in (-1,0)$,
\end{enumerate}
and four different specifications of $\eta(t)$
\begin{enumerate}
\item Independent and identically distributed (IID) Gaussian innovations: In this setting, we specify
\begin{equation}
\eta(t)  \sim \mathscr{N}(0,1)
\label{Setting:1}
\end{equation}
as IID random numbers drawn from the standard Gaussian distribution.

\item IID shifted Gamma innovation: In this setting,
we specify
\begin{equation}
\eta(t) = \gamma_0(t) - \sqrt{2}
\label{Setting:2}
\end{equation}
where
$\left\{\gamma_0(t)\right\}_{t=0}^{d-1}$ were random numbers drawn randomly from Gamma distribution with shape parameter 2 and scale parameter $2^{-1/2}$, i.e.,
$\Gamma(2,2^{-1/2})$.

\item IID scaled pseuo-uniform innovation: In this setting, we specify
\begin{equation}
\eta(t) = \sqrt{12} \left( \exp( t + 3 \arctan (s)
- \lfloor  \exp(t + 3 \arctan (s))  \rfloor - 1/2 \right)
\label{Setting:4}
\end{equation}
where $t = 1,\ldots,d-1$,  $s = 1,\ldots,N_{sim},$
and where $N_{sim}$ is the total number of simulations to be performed.

\item Auto-correlated Gaussian innovation: In this setting, we specify
\begin{equation}
\eta(t) = 2^{-1/2} \left( \eta_0(t) +  \eta_0(t-1) \right)
\label{Setting:3}
\end{equation}
where
$\left\{\eta_0(t)\right\}_{t=0}^{d-1}$ were IID random numbers drawn randomly from $\mathscr{N}(0,1)$ and the lag-one auto-correlation is
$\E[\eta(t) \eta(t-1)] = 0.5$.

\end{enumerate}

For this simulation experiment,
the linear predictors
\eqref{prediction:kernel:predictor}
and
\eqref{smooth:and:prediction:kernel:predictor}
are implemented in the form of \eqref{y(t):sequence}
as
\begin{equation}
y(d-1) = \sum_{\tau =0}^{d-1}
h(t-\tau) \; x(\tau) \approx x(d) \;,
\label{y(d)}
\end{equation}
where
$y(d-1) = y_{\K \H}(d-1)$ and
$y(d-1) = y_{\K}(d-1)$ are the one-step-ahead forecasts,
and
where
$h(t-\tau)=h_{\K \H}(t-\tau)$ and
$h(t-\tau)=h_{\K}(t-\tau)$
are the impulse response  functions for
\eqref{prediction:kernel:predictor}
and
\eqref{smooth:and:prediction:kernel:predictor}
respectively.
Following the choice of
parameters used in \cite{Dokuchaev:SP:2016},
we set
$a = 0.6, p = 0.7, N = 100, m = 2$,
for the smoothing filter \eqref{smooth:transfer:function},
and
set
$\gamma = 1.1$,
for the predicting kernel \eqref{predictor:transfer:function}.
We investigate the sensitivity of the predicting kernel for some different values of $r$,
where
$r \in \{0.8, 1.1, 1.5, 2 \}$,
and
we consider three different lengths of ultra short  sequence $d$,
where
$d \in \left\{ 4, 5, 6 \right\}$. \par

The ideal linear predictor
\eqref{optimal:predictor}
is implemented as
\begin{equation}
y_{\scriptscriptstyle ideal}(d-1)
=
\beta_2(d) \; x(d-1) \approx x(d) \;.
\label{sample:optimal:predictor}
\end{equation}
For this predictor, one needs to know $\beta_2(d)$, i.e., the post-break AR(1) model parameters used to simulate $x(d)$.
In practice, it is impossible to know $\beta_2(d)$.
We include \eqref{sample:optimal:predictor} as it represents a theoretical ideal benchmark. \par

The linear predictor
\eqref{AR1:predictor}
is implemented as
\begin{equation}
y_{\scriptscriptstyle AR(1)}(d)
=
\hat{\beta}(d) \; x(d-1) \approx x(d) \;,
\label{sample:linear:predictor}
\end{equation}
where
$\hat{\beta}(d)$
is estimated
by fitting an AR(1) model to the learning sequence
$\left\{ x(\tau) \right\}_{\tau=0}^{d-1}$.
This is a commonly used approach in AR(1) time series forecasting, and one that depends on the large-sample properties of the available time series for efficient model parameter estimation. We are interested to investigate the finite-sample properties in terms of  forecast performance of \eqref{y(d)}
relative to \eqref{sample:linear:predictor} in the context of ultra short learning sequences considered in this paper. \par

For each combination of
$(\beta_1, \beta_2)$,
$\eta(t)$,
$r$,
and
$d$,
we perform $N_{sim}$ simulations
where $N_{sim} \in \{ 1 \times 10^5, 2 \times 10^5, 3 \times 10^5 \}$.
For each simulation,
we simulate an
AR(1) processes with a single random structural break at random unknown time point following the data generation process \eqref{x(t):sequence},
each containing $d+1$ observations.
where $\beta_1(t)$, $\beta_2(t)$, $\sigma$, and $\eta(t)$
are mutually independent.
The first $d$ observations are used as the sequence based on which we forecast the $(d+1)$-th observation.
 \par

Let $\mathbb{E}[\cdot]$ denote the sample mean across the
$N_{sim}$ Monte Carlo trials performed for each scenario indexed by $s$ where $s=1,\ldots,N_{sim}$.
Specifically, we let
\begin{align*}
e_{\K \H}
&=
\left( \;
\mathbb{E} \left( x(d) - y_{\K \H}(d-1)  \right)^2 \; \right)^{1/2} \;, \\
e_{\K}
&=
\left( \;
\mathbb{E} \left( x(d) - y_{\K}(d-1)  \right)^2 \; \right)^{1/2} \;, \\
e_{\scriptscriptstyle ideal}
&=
\left( \;
\mathbb{E} \left( x(d) - y_{\scriptscriptstyle ideal}(d-1)  \right)^2 \; \right)^{1/2} \;, \\
e_{\scriptscriptstyle AR(1)}
&=
\left( \;
\mathbb{E} \left( x(d) - y_{\scriptscriptstyle AR(1)}(d-1)  \right)^2 \; \right)^{1/2} \;,
\end{align*}
be the sample root-mean-squared error (RMSE) for
\eqref{y(d)}
that implement
\eqref{prediction:kernel:predictor}
and
\eqref{smooth:and:prediction:kernel:predictor},
\eqref{sample:optimal:predictor},
and
\eqref{sample:linear:predictor}
respectively.

We carry out the simulation experiments in the {\tt R}
computing environment
\citep{R:2016}.
Simulation of the learning sequence is carried out
by iterative application of
\eqref{x(t):sequence}.
The estimation of AR(1) parameter $\hat{\beta}$
for the implementation of  \eqref{sample:linear:predictor}
is performed using the {\tt ar.ols()} script in {\tt R}.
Numerical integrations
carried out to map
\eqref{prediction:kernel:predictor}
and
\eqref{smooth:and:prediction:kernel:predictor}
to their respective impulse response functions
to be used in
\eqref{y(d)}
are implemented via the {\tt myintegrate()} script in the
{\tt R}
add-on package {\tt elliptic} proposed in
\cite{Hankin:2006}.
 \par

\begin{table}[h]
\centering
\caption{One-step-ahead forecast performance with
$\beta_1, \beta_2 \in (0,1)$,
random structural break at $\theta \in \{ 2, \ldots, d-2 \}$,  and
$\eta(t) \sim \mathscr{N}(0,1)$.}
\label{Simulation:results:gaussian:noise:positive:both:beta:random:break}
\scalebox{0.8}{
\begin{tabular}{lcccccccc}
  \hline
&
&$e_{\scriptscriptstyle ideal}$
&$e_{\scriptscriptstyle AR(1)}$
&$e_{\K}$
&$e_{\K \H}$
&$e_{\scriptscriptstyle ideal}  /  e_{\scriptscriptstyle AR(1)}$
&$e_{\K}  /  e_{\scriptscriptstyle AR(1)}$
 &$e_{\K \H}  /  e_{\scriptscriptstyle AR(1)}$ \\
  \hline \\

&\multicolumn{8}{c}{Panel (a): $\beta_1, \beta_2 \in (0,1), N_{sim}=1 \times 10^5$} \\

$r = 0.8, d=4$   && 0.30031 & 0.44385 & 0.39284 & 0.34568 & 0.67661 & 0.88506 & 0.77882   \\
$r = 0.8, d=5$   && 0.29881 & 0.37738 & 0.39343 & 0.34661 & 0.79182 & 1.04254 & 0.91847   \\
$r = 0.8, d=6$   && 0.29970 & 0.35256 & 0.39499 & 0.34795 & 0.85004 & 1.12032 & 0.98691    \\  \\

$r = 1.1, d=4$   && 0.30082 & 0.44232 & 0.39817 & 0.34568 & 0.68010 & 0.90018 & 0.78152   \\
$r = 1.1, d=5$   && 0.29971 & 0.37597 & 0.39765 & 0.34586 & 0.79717 & 1.05767 & 0.91991   \\
$r = 1.1, d=6$   && 0.29986 & 0.35161 & 0.39970 & 0.34686 & 0.85283 & 1.13678 & 0.98640    \\  \\

$r = 1.5, d=4$   && 0.30066 & 0.46608 & 0.40348 & 0.34393 & 0.64508 & 0.86568 & 0.73791   \\
$r = 1.5, d=5$   && 0.29979 & 0.37684 & 0.40480 & 0.34393 & 0.79554 & 1.07420 & 0.91267   \\
$r = 1.5, d=6$   && 0.29939 & 0.35283 & 0.40506 & 0.34468 & 0.84853 & 1.14802 & 0.97689    \\  \\

$r = 2, d=4$   && 0.30010 & 0.44740 & 0.41242 & 0.34097 & 0.67077 & 0.92182 & 0.76210   \\
$r = 2, d=5$   && 0.30127 & 0.37954 & 0.41351 & 0.34376 & 0.79379 & 1.08950 & 0.90572   \\
$r = 2, d=6$   && 0.30108 & 0.35406 & 0.41589 & 0.34351 & 0.85037 & 1.17464 & 0.97022     \\  \\

&\multicolumn{8}{c}{Panel (b): $\beta_1, \beta_2 \in (0,1), N_{sim}=2 \times 10^5$} \\

$r = 0.8, d=4$   && 0.29981 & 0.44456 & 0.39188 & 0.34614 & 0.67439 & 0.88152 & 0.77862   \\
$r = 0.8, d=5$   && 0.29951 & 0.37979 & 0.39302 & 0.34690 & 0.78861 & 1.03484 & 0.91340   \\
$r = 0.8, d=6$   && 0.30024 & 0.35278 & 0.39404 & 0.34810 & 0.85107 & 1.11696 & 0.98673    \\  \\

$r = 1.1, d=4$   && 0.29946 & 0.44962 & 0.39626 & 0.34470 & 0.66603 & 0.88132 & 0.76666   \\
$r = 1.1, d=5$   && 0.29959 & 0.38039 & 0.39761 & 0.34553 & 0.78759 & 1.04526 & 0.90836    \\
$r = 1.1, d=6$   && 0.29981 & 0.35372 & 0.39946 & 0.34672 & 0.84759 & 1.12934 & 0.98021    \\  \\

$r = 1.5, d=4$   && 0.30063 & 0.44241 & 0.40416 & 0.34362 & 0.67954 & 0.91355 & 0.77671   \\
$r = 1.5, d=5$   && 0.30073 & 0.38054 & 0.40539 & 0.34494 & 0.79027 & 1.06529 & 0.90643   \\
$r = 1.5, d=6$   && 0.29988 & 0.35258 & 0.40506 & 0.34588 & 0.85053 & 1.14885 & 0.98100    \\  \\

$r = 2, d=4$   && 0.30021 & 0.44961 & 0.41343 & 0.34147 & 0.66770 & 0.91953 & 0.75948   \\
$r = 2, d=5$   && 0.29953 & 0.37829 & 0.41324 & 0.34205 & 0.79179 & 1.09237 & 0.90420   \\
$r = 2, d=6$   && 0.30112 & 0.35523 & 0.41516 & 0.34450 & 0.84766 & 1.16869 & 0.96978     \\  \\

&\multicolumn{8}{c}{Panel (c): $\beta_1, \beta_2 \in (0,1), N_{sim}=3 \times 10^5$} \\

$r = 0.8, d=4$   && 0.30000 & 0.44340 & 0.39318 & 0.34638 & 0.67659 & 0.88675 & 0.78119   \\
$r = 0.8, d=5$   && 0.29977 & 0.38060 & 0.39391 & 0.34738 & 0.78762 & 1.03496 & 0.91271   \\
$r = 0.8, d=6$   && 0.30012 & 0.35344 & 0.39397 & 0.34852 & 0.84913 & 1.11468 & 0.98607    \\  \\

$r = 1.1, d=4$   && 0.30019 & 0.44827 & 0.39724 & 0.34511 & 0.66967 & 0.88617 & 0.76988  \\
$r = 1.1, d=5$   && 0.30003 & 0.37772 & 0.39832 & 0.34573 & 0.79431 & 1.05454 & 0.91530   \\
$r = 1.1, d=6$   && 0.30053 & 0.35391 & 0.39903 & 0.34743 & 0.84917 & 1.12750 & 0.98171   \\  \\

$r = 1.5, d=4$   && 0.30025 & 0.44945 & 0.40429 & 0.34404 & 0.66804 & 0.89953 & 0.76547   \\
$r = 1.5, d=5$   && 0.29953 & 0.37719 & 0.40392 & 0.34392 & 0.79412 & 1.07088 & 0.91180   \\
$r = 1.5, d=6$   && 0.29961 & 0.35373 & 0.40467 & 0.34501 & 0.84700 & 1.14402 & 0.97537    \\  \\

$r = 2, d=4$   && 0.30004 & 0.44259 & 0.41150 & 0.34183 & 0.67793 & 0.92975 & 0.77233   \\
$r = 2, d=5$   && 0.29985 & 0.37641 & 0.41410 & 0.34193 & 0.79660 & 1.10013 & 0.90838   \\
$r = 2, d=6$   && 0.29997 & 0.35304 & 0.41438 & 0.34299 & 0.84967 & 1.17374 & 0.97152     \\  \\

   \hline
\end{tabular}
}

\end{table}

Table \ref{Simulation:results:gaussian:noise:positive:both:beta:random:break} depicts the simulation experiments results
for the setting with
$\beta_1, \beta_2 \in (0,1)$
and
$\eta(t) \sim \mathscr{N}(0,1)$.
For a short sequence of length $d=4, 5, 6$,
and
for the four different values of predicting kernel parameter $r$,
$r=0.8, 1.1, 1.5, 2$,
the RMSE of
the smoothed predicting kernel linear predictor,
is smaller than
the RMSE of the linear predictor that utilize
AR(1) model parameter
estimated based on the learning sequence
ignoring the presence of structural break
\eqref{sample:linear:predictor}.
The shorter the learning sequence, the
better the performance of the smoothed predicting kernel linear predictor.
This trend is consistent across three different sizes of Monte Carlo simulation
$N_{sim} \in \{ 1 \times 10^4, 2 \times 10^4, 3 \times 10^4 \}$. \par

Worthy of note is that
this smoothed predictor does not require explicit modelling of structural break.
In practice,
when the available learning sequence is short, and the model parameter structural break time and magnitude uncertain, it is not possible to efficiently apply structural break estimation and adjustment procedures for parameter estimation and time series forecasting due to series length constraint.
In this context,
the smoothed predicting kernel
\eqref{smooth:and:prediction:kernel:predictor}
appears to be an alternative approach that may offer satisfactory forecast performance,
circumventing the need of
resorting to model parameter estimation that ignore structure break. \par

The fact that
the RMSE of the predicting kernel linear predictor without smoothing
\eqref{prediction:kernel:predictor}
is larger than
the RMSE of the linear predictor
\eqref{sample:linear:predictor}
highlights the
role of the near-ideal causal smoother
\eqref{smooth:transfer:function}
in improving the forecast performance of
\eqref{prediction:kernel:predictor}.
By dampening the high frequency noise,
the smoothed prediction kernel is more able to capture the salient features of the simulated AR(1) process with random structural break
based on a short learning sequence
in order to deliver better
one-step-ahead forecast performance
than
the linear predictor
\eqref{sample:linear:predictor}.
Without the aid of the smoothing kernel,
the performance of the predicting kernel
\eqref{prediction:kernel:predictor}
is, in general, even poorer than that of the linear predictor
\eqref{sample:linear:predictor}
that relies on model parameter estimate from an AR(1) model that ignores structural break. \par

It is not surprising that the performance of the linear predictor
\eqref{sample:linear:predictor}
is poorer than the
ideal predictor
\eqref{sample:optimal:predictor}.
Utilizing pre-break and post-break data to estimate post-break model parameter when the break time and magnitude are unknown
inevitably leads to parameter estimation error.
Although
cross-validation methods have been proposed
to utilize pre-break and post-break data
to use pre-break data to estimate the parameters of the model used to compute out-of sample
forecasts
\citep[see, e.g.,][]{Pesaran:Timmermann:2007},
the number of observations required to implement these methodologies is considerably larger than
those we consider in this paper. \par

Table \ref{Simulation:results:gaussian:noise:other:settings} depicts the results of simulations with $\eta(t) \sim \mathscr{N}(0,1)$ for
three other model parameter settings, i.e.,
a wider range of possible model parameters with $\beta_1, \beta_2 \in (-1,1)$,
a model parameter shift from negative to positive values
$\beta_1 \in (-1,0), \beta_2 \in (0,1)$,
and
a model parameter shift from positive to negative values
$\beta_1 \in (0,1), \beta_2 \in (-1,0)$.
The forecast performance of
the smoothed predicting kernel linear predictor
\eqref{smooth:and:prediction:kernel:predictor}
appears to be dependent on the sign of $\beta_2$.
If $\beta_2 \in (0,1)$,
\eqref{smooth:and:prediction:kernel:predictor}
performs better than
\eqref{sample:linear:predictor}, and vice versa.
\par

Table \ref{Simulation:results:Gamma:noise:all:settings}
depicts
a subset of the
simulation results pertaining to the setting where $\eta(t)$
is as defined in \eqref{Setting:2},
while
Table \ref{Simulation:results:pseudo:random:noise:all:settings}
depicts
those pertaining to the setting where $\eta(t)$
is as defined in \eqref{Setting:3}.
They shows
similar trends as those demonstrated for
those of \eqref{Setting:1} as depicted in
Table \ref{Simulation:results:gaussian:noise:positive:both:beta:random:break}
and
Table \ref{Simulation:results:gaussian:noise:other:settings}
above.  \par

However, the numerical results pertaining to simulation scenarios with IID innovation terms
are
in some ways different from those with correlated innovation terms.
Table \ref{Simulation:results:AR1:noise:all:settings}
depicts
a subset of the
simulation results pertaining to the setting where $\eta(t)$
is as defined in \eqref{Setting:4}
where $\E[\eta(t) \eta(t-1)] = 0.5$.
While the forecast performance of the smoothed predicting kernel
\eqref{smooth:and:prediction:kernel:predictor}
is still better than
the linear predictor
\eqref{sample:linear:predictor}
for $\beta_1 \in (-1,0)$ and $\beta_2 \in (0,1)$
in this context where the innovation terms are auto-correlated,
it is not so
for the remaining three simulation scenarios. \par

\begin{table}[t]
\centering
\caption{ Random structural break at $\theta \in \{ 2, \ldots, d-2 \}$, $\eta(t) \sim \mathscr{N}(0,1)$ and $N_{sim} = 3 \times 10^5$.}
\label{Simulation:results:gaussian:noise:other:settings}
\scalebox{0.8}{
\begin{tabular}{lcccccccc}
  \hline
&
&$e_{\scriptscriptstyle ideal}$
&$e_{\scriptscriptstyle AR(1)}$
&$e_{\K}$
&$e_{\K \H}$
&$e_{\scriptscriptstyle ideal}  /  e_{\scriptscriptstyle AR(1)}$
&$e_{\K}  /  e_{\scriptscriptstyle AR(1)}$
 &$e_{\K \H}  /  e_{\scriptscriptstyle AR(1)}$ \\
  \hline \\

  &\multicolumn{8}{c}{Panel (a): $\beta_1, \beta_2 \in (-1,1)$} \\

$r = 0.8, d=4$   && 0.29946 & 0.44548 & 0.72830 & 0.42322 & 0.67221 & 1.63486 & 0.95002   \\
$r = 0.8, d=5$   && 0.29988 & 0.38167 & 0.75529 & 0.42575 & 0.78570 & 1.97890 & 1.11548   \\
$r = 0.8, d=6$   && 0.29970 & 0.36081 & 0.77564 & 0.42754 & 0.83064 & 2.14972 & 1.18496   \\  \\

$r = 1.1, d=4$   && 0.29967 & 0.44583 & 0.73269 & 0.42220 & 0.67215 & 1.64341 & 0.94699   \\
$r = 1.1, d=5$   && 0.30002 & 0.38609 & 0.76609 & 0.42452 & 0.77705 & 1.98421 & 1.09952   \\
$r = 1.1, d=6$   && 0.29954 & 0.36060 & 0.79259 & 0.42832 & 0.83065 & 2.19794 & 1.18778    \\  \\

$r = 1.5, d=4$   && 0.29981 & 0.45026 & 0.75434 & 0.42250 & 0.66587 & 1.67533 & 0.93834    \\
$r = 1.5, d=5$   && 0.30058 & 0.38192 & 0.78852 & 0.42577 & 0.78702 & 2.06463 & 1.11482   \\
$r = 1.5, d=6$   && 0.29974 & 0.36132 & 0.81562 & 0.42766 & 0.82957 & 2.25735 & 1.18361    \\  \\

$r = 2, d=4$   && 0.29950 & 0.45461 & 0.77642 & 0.42232 & 0.65880 & 1.70790 & 0.92899   \\
$r = 2, d=5$   && 0.29958 & 0.38185 & 0.81207 & 0.42434 & 0.78454 & 2.12665 & 1.11127   \\
$r = 2, d=6$   && 0.29965 & 0.36041 & 0.84057 & 0.42723 & 0.83141 & 2.33228 & 1.18541    \\  \\

  &\multicolumn{8}{c}{Panel (b): $\beta_1 \in (-1,0), \beta_2 \in (0,1)$} \\

$r = 0.8, d=4$   && 0.30020 & 0.44990 & 0.39408 & 0.34707 & 0.66727 & 0.87593 & 0.77143   \\
$r = 0.8, d=5$   && 0.30020 & 0.44990 & 0.39408 & 0.34707 & 0.66727 & 0.87593 & 0.77143   \\
$r = 0.8, d=6$   && 0.29935 & 0.36745 & 0.39601 & 0.34919 & 0.81467 & 1.07770 & 0.95028     \\  \\

$r = 1.1, d=4$   && 0.29977 & 0.44824 & 0.39914 & 0.34585 & 0.66876 & 0.89047 & 0.77157   \\
$r = 1.1, d=5$   && 0.29969 & 0.38747 & 0.40141 & 0.34654 & 0.77346 & 1.03598 & 0.89435   \\
$r = 1.1, d=6$   && 0.30021 & 0.36857 & 0.40261 & 0.34920 & 0.81455 & 1.09238 & 0.94745    \\  \\

$r = 1.5, d=4$   && 0.30021 & 0.44596 & 0.40667 & 0.34529 & 0.67318 & 0.91189 & 0.77426   \\
$r = 1.5, d=5$   && 0.29937 & 0.38618 & 0.40881 & 0.34472 & 0.77521 & 1.05861 & 0.89264  \\
$r = 1.5, d=6$   && 0.30039 & 0.36888 & 0.41055 & 0.34667 & 0.81433 & 1.11297 & 0.93979    \\  \\

$r = 2, d=4$   && 0.30062 & 0.47598 & 0.41782 & 0.34428 & 0.63158 & 0.87781 & 0.72331    \\
$r = 2, d=5$   && 0.30000 & 0.38460 & 0.42137 & 0.34384 & 0.78005 & 1.09562 & 0.89404    \\
$r = 2, d=6$   && 0.29983 & 0.36798 & 0.42229 & 0.34552 & 0.81481 & 1.14760 & 0.93896    \\  \\

  &\multicolumn{8}{c}{Panel (c): $\beta_1 \in (0,1), \beta_2 \in (-1,0)$} \\

$r = 0.8, d=4$   && 0.29982 & 0.44890 & 0.94541 & 0.48679 & 0.66790 & 2.10604 & 1.08439   \\
$r = 0.8, d=5$   && 0.29966 & 0.38576 & 0.98889 & 0.49084 & 0.77681 & 2.56348 & 1.27239   \\
$r = 0.8, d=6$   && 0.30018 & 0.36923 & 1.01892 & 0.49363 & 0.81300 & 2.75955 & 1.33692    \\  \\

$r = 1.1, d=4$   && 0.30018 & 0.45163 & 0.95963 & 0.48872 & 0.66466 & 2.12480 & 1.08212   \\
$r = 1.1, d=5$   && 0.29999 & 0.38558 & 1.00126 & 0.49075 & 0.77802 & 2.59674 & 1.27275   \\
$r = 1.1, d=6$   && 0.29929 & 0.36789 & 1.03947 & 0.49352 & 0.81353 & 2.82552 & 1.34151    \\  \\

$r = 1.5, d=4$   && 0.30073 & 0.44550 & 0.97856 & 0.48733 & 0.67504 & 2.19655 & 1.09391   \\
$r = 1.5, d=5$   && 0.29968 & 0.38577 & 1.02990 & 0.49081 & 0.77683 & 2.66972 & 1.27229   \\
$r = 1.5, d=6$   && 0.29997 & 0.36922 & 1.07103 & 0.49593 & 0.81244 & 2.90078 & 1.34319     \\  \\

$r = 2, d=4$   && 0.29985 & 0.45724 & 1.00339 & 0.48925 & 0.65579 & 2.19445 & 1.07001   \\
$r = 2, d=5$   && 0.29995 & 0.38617 & 1.05628 & 0.49223 & 0.77673 & 2.73525 & 1.27464   \\
$r = 2, d=6$   && 0.29945 & 0.36725 & 1.10925 & 0.49614 & 0.81538 & 3.02041 & 1.35097    \\  \\

   \hline
\end{tabular}
}

\end{table}

\begin{table}[t]
\centering
\caption{One-step-ahead forecast performance with random structural break at $\theta \in \{ 2, \ldots, d-2 \}$,  and
$\eta(t) = \gamma_0(t) - \sqrt{2}$, where $\gamma_0(t) \sim \Gamma(2,2^{-1/2})$, and $N_{sim}=3 \times 10^5$.}
\label{Simulation:results:Gamma:noise:all:settings}
\scalebox{0.9}{
\begin{tabular}{lcccccccc}
  \hline
&
&$e_{\scriptscriptstyle ideal}$
&$e_{\scriptscriptstyle AR(1)}$
&$e_{\K}$
&$e_{\K \H}$
&$e_{\scriptscriptstyle ideal}  /  e_{\scriptscriptstyle AR(1)}$
&$e_{\K}  /  e_{\scriptscriptstyle AR(1)}$
 &$e_{\K \H}  /  e_{\scriptscriptstyle AR(1)}$ \\
  \hline \\

&\multicolumn{8}{c}{Panel (a): $\beta_1, \beta_2 \in (0,1)$} \\

$r = 0.8, d=4$   && 0.30084 & 0.53991 & 0.39310 & 0.34721 & 0.55721 & 0.72808 & 0.64308    \\
$r = 0.8, d=5$   && 0.30016 & 0.42423 & 0.39442 & 0.34722 & 0.70754 & 0.92973 & 0.81846   \\
$r = 0.8, d=6$   && 0.29990 & 0.37654 & 0.39329 & 0.34776 & 0.79645 & 1.04447 & 0.92357     \\  \\

$r = 2, d=4$   && 0.29977 & 0.52893 & 0.41234 & 0.34127 & 0.56676 & 0.77957 & 0.64520   \\
$r = 2, d=5$   && 0.30001 & 0.41651 & 0.41255 & 0.34243 & 0.72029 & 0.99049 & 0.82214   \\
$r = 2, d=6$   && 0.30028 & 0.37886 & 0.41402 & 0.34331 & 0.79259 & 1.09281 & 0.90616    \\  \\

  &\multicolumn{8}{c}{Panel (b): $\beta_1, \beta_2 \in (-1,1)$} \\

$r = 0.8, d=4$   && 0.29997 & 0.56377 & 0.72394 & 0.42278 & 0.53207 & 1.28410 & 0.74992  \\
$r = 0.8, d=5$   && 0.29972 & 0.42580 & 0.75337 & 0.42486 & 0.70390 & 1.76931 & 0.99780   \\
$r = 0.8, d=6$   && 0.29922 & 0.38669 & 0.78082 & 0.42880 & 0.77379 & 2.01922 & 1.10888   \\  \\

$r = 2, d=4$   && 0.29970 & 0.53578 & 0.77402 & 0.42328 & 0.55937 & 1.44466 & 0.79002  \\
$r = 2, d=5$   && 0.30021 & 0.42405 & 0.81194 & 0.42479 & 0.70796 & 1.91473 & 1.00174   \\
$r = 2, d=6$   && 0.30050 & 0.38871 & 0.83975 & 0.42698 & 0.77309 & 2.16039 & 1.09847     \\  \\

  &\multicolumn{8}{c}{Panel (c): $\beta_1 \in (-1,0), \beta_2 \in (0,1)$} \\

$r = 0.8, d=4$   && 0.29942 & 0.53158 & 0.39353 & 0.34666 & 0.56327 & 0.74029 & 0.65212   \\
$r = 0.8, d=5$   && 0.29993 & 0.43544 & 0.39468 & 0.34786 & 0.68881 & 0.90640 & 0.79888   \\
$r = 0.8, d=6$   && 0.30003 & 0.39344 & 0.39514 & 0.34948 & 0.76257 & 1.00431 & 0.88827     \\  \\

$r = 2, d=4$   && 0.30088 & 0.53712 & 0.41771 & 0.34455 & 0.56018 & 0.77770 & 0.64149   \\
$r = 2, d=5$   && 0.29959 & 0.43570 & 0.42138 & 0.34349 & 0.68760 & 0.96712 & 0.78835    \\
$r = 2, d=6$   && 0.29957 & 0.39110 & 0.42320 & 0.34473 & 0.76595 & 1.08208 & 0.88144    \\  \\

  &\multicolumn{8}{c}{Panel (d): $\beta_1 \in (0,1), \beta_2 \in (-1,0)$} \\

$r = 0.8, d=4$   && 0.29991 & 0.54170 & 0.94341 & 0.48621 & 0.55365 & 1.74155 & 0.89756   \\
$r = 0.8, d=5$   && 0.30044 & 0.42537 & 0.98715 & 0.49148 & 0.70630 & 2.32067 & 1.15542   \\
$r = 0.8, d=6$   && 0.30079 & 0.39497 & 1.02682 & 0.49603 & 0.76155 & 2.59973 & 1.25588    \\  \\

$r = 2, d=4$   && 0.30080 & 0.52253 & 1.00837 & 0.49099 & 0.57565 & 1.92978 & 0.93963    \\
$r = 2, d=5$   && 0.29956 & 0.43413 & 1.06041 & 0.49239 & 0.69004 & 2.44263 & 1.13420  \\
$r = 2, d=6$   && 0.30043 & 0.39440 & 1.10692 & 0.49708 & 0.76174 & 2.80658 & 1.26033    \\  \\

   \hline
\end{tabular}
}

\end{table}

\begin{table}[t]
\centering
\caption{One-step-ahead forecast performance with random structural break at $\theta \in \{ 2, \ldots, d-2 \}$,
$\eta(t) = \sqrt{12} \left( \exp( t + 3 \arctan (s) - \lfloor  \exp(t + 3 \arctan (s))  \rfloor - 1/2 \right)$, and  $N_{sim}=3 \times 10^5$.}
\label{Simulation:results:pseudo:random:noise:all:settings}
\scalebox{0.9}{
\begin{tabular}{lcccccccc}
  \hline
&
&$e_{\scriptscriptstyle ideal}$
&$e_{\scriptscriptstyle AR(1)}$
&$e_{\K}$
&$e_{\K \H}$
&$e_{\scriptscriptstyle ideal}  /  e_{\scriptscriptstyle AR(1)}$
&$e_{\K}  /  e_{\scriptscriptstyle AR(1)}$
 &$e_{\K \H}  /  e_{\scriptscriptstyle AR(1)}$ \\
  \hline \\

&\multicolumn{8}{c}{Panel (a): $\beta_1, \beta_2 \in (0,1)$} \\

$r = 0.8, d=4$   && 0.27537 & 0.30305 & 0.33968 & 0.22866 & 0.90865 & 1.12087 & 0.75455   \\
$r = 0.8, d=5$   && 0.31588 & 0.49982 & 0.39241 & 0.35066 & 0.63198 & 0.78510 & 0.70157    \\
$r = 0.8, d=6$   && 0.28991 & 0.36228 & 0.39589 & 0.32990 & 0.80024 & 1.09278 & 0.91063    \\  \\

$r = 2, d=4$   && 0.27537 & 0.30289 & 0.37362 & 0.23235 & 0.90912 & 1.23352 & 0.76710    \\
$r = 2, d=5$   && 0.31588 & 0.49950 & 0.40473 & 0.34447 & 0.63239 & 0.81027 & 0.68963   \\
$r = 2, d=6$   && 0.28991 & 0.36230 & 0.41445 & 0.32531 & 0.80019 & 1.14393 & 0.89790    \\  \\

  &\multicolumn{8}{c}{Panel (b): $\beta_1, \beta_2 \in (-1,1)$} \\

$r = 0.8, d=4$   && 0.27537 & 0.33724 & 0.52333 & 0.25020 & 0.81653 & 1.55182 & 0.74190   \\
$r = 0.8, d=5$   && 0.31588 & 0.45315 & 0.66214 & 0.41907 & 0.69707 & 1.46117 & 0.92478   \\
$r = 0.8, d=6$   && 0.28991 & 0.35271 & 0.71703 & 0.42148 & 0.82194 & 2.03290 & 1.19495    \\  \\

$r = 2, d=4$   && 0.27537 & 0.33766 & 0.56239 & 0.25452 & 0.81551 & 1.66555 & 0.75377   \\
$r = 2, d=5$   && 0.31588 & 0.45277 & 0.70309 & 0.41578 & 0.69766 & 1.55286 & 0.91829   \\
$r = 2, d=6$   && 0.28991 & 0.35254 & 0.76555 & 0.42132 & 0.82235 & 2.17153 & 1.19508    \\  \\

  &\multicolumn{8}{c}{Panel (c): $\beta_1 \in (-1,0), \beta_2 \in (0,1)$} \\

$r = 0.8, d=4$   && 0.27537 & 0.30559 & 0.34224 & 0.22801 & 0.90109 & 1.11993 & 0.74613   \\
$r = 0.8, d=5$   && 0.31588 & 0.48764 & 0.39469 & 0.35113 & 0.64777 & 0.80937 & 0.72005   \\
$r = 0.8, d=6$   && 0.28991 & 0.35639 & 0.39934 & 0.33230 & 0.81346 & 1.12052 & 0.93241     \\  \\

$r = 2, d=4$   && 0.27537 & 0.30552 & 0.37980 & 0.23065 & 0.90129 & 1.24313 & 0.75495   \\
$r = 2, d=5$   && 0.31588 & 0.48764 & 0.40948 & 0.34395 & 0.64778 & 0.83973 & 0.70535   \\
$r = 2, d=6$   && 0.28991 & 0.35629 & 0.42379 & 0.32902 & 0.81369 & 1.18945 & 0.92346    \\  \\

  &\multicolumn{8}{c}{Panel (d): $\beta_1 \in (0,1), \beta_2 \in (-1,0)$} \\

$r = 0.8, d=4$   && 0.27537 & 0.37214 & 0.62981 & 0.26768 & 0.73995 & 1.69240 & 0.71930   \\
$r = 0.8, d=5$   && 0.31588 & 0.42228 & 0.83730 & 0.47762 & 0.74804 & 1.98282 & 1.13106    \\
$r = 0.8, d=6$   && 0.28991 & 0.35600 & 0.90941 & 0.49547 & 0.81435 & 2.55451 & 1.39176    \\  \\

$r = 2, d=4$   && 0.27537 & 0.37196 & 0.67132 & 0.27473 & 0.74032 & 1.80483 & 0.73860   \\
$r = 2, d=5$   && 0.31588 & 0.42222 & 0.89167 & 0.47730 & 0.74815 & 2.11187 & 1.13045   \\
$r = 2, d=6$   && 0.28991 & 0.35638 & 0.96375 & 0.49639 & 0.81349 & 2.70429 & 1.39286    \\  \\

   \hline
\end{tabular}
}

\end{table}

\begin{table}[t]
\centering
\caption{One-step-ahead forecast performance with random structural break at $\theta \in \{ 2, \ldots, d-2 \}$,  and
$\theta \in \{ 2, \ldots, d-2 \}$, $\eta(t) = 2^{-1/2} \eta_0(t) + 2^{-1/2} \eta_0(t-1)$, where
$\eta_0(t) \sim \mathscr{N}(0,1)$, and  $N_{sim}=3 \times 10^5$.}
\label{Simulation:results:AR1:noise:all:settings}
\scalebox{0.9}{
\begin{tabular}{lcccccccc}
  \hline
&
&$e_{\scriptscriptstyle ideal}$
&$e_{\scriptscriptstyle AR(1)}$
&$e_{\K}$
&$e_{\K \H}$
&$e_{\scriptscriptstyle ideal}  /  e_{\scriptscriptstyle AR(1)}$
&$e_{\K}  /  e_{\scriptscriptstyle AR(1)}$
 &$e_{\K \H}  /  e_{\scriptscriptstyle AR(1)}$ \\
  \hline \\

&\multicolumn{8}{c}{Panel (a): $\beta_1, \beta_2 \in (0,1)$} \\

$r = 0.8, d=4$   && 0.30035 & 0.41511 & 0.26556 & 0.33784 & 0.72355 & 0.63973 & 0.81386   \\
$r = 0.8, d=5$   && 0.30030 & 0.35935 & 0.26788 & 0.33906 & 0.83566 & 0.74544 & 0.94354    \\
$r = 0.8, d=6$   && 0.29979 & 0.33688 & 0.26966 & 0.33870 & 0.88989 & 0.80046 & 1.00540    \\  \\

$r = 2, d=4$   && 0.30093 & 0.41192 & 0.26972 & 0.33210 & 0.73057 & 0.65479 & 0.80623   \\
$r = 2, d=5$   && 0.29996 & 0.36100 & 0.27102 & 0.33213 & 0.83093 & 0.75077 & 0.92005   \\
$r = 2, d=6$   && 0.30007 & 0.33697 & 0.27332 & 0.33336 & 0.89049 & 0.81111 & 0.98927     \\  \\

  &\multicolumn{8}{c}{Panel (b): $\beta_1, \beta_2 \in (-1,1)$} \\

$r = 0.8, d=4$   && 0.29964 & 0.38414 & 0.36390 & 0.32763 & 0.78002 & 0.94730 & 0.85290   \\
$r = 0.8, d=5$   && 0.29966 & 0.33359 & 0.37153 & 0.32864 & 0.89830 & 1.11373 & 0.98518   \\
$r = 0.8, d=6$   && 0.30046 & 0.31771 & 0.37667 & 0.33065 & 0.94571 & 1.18558 & 1.04076    \\  \\

$r = 2, d=4$   && 0.30027 & 0.38045 & 0.38446 & 0.32176 & 0.78925 & 1.01055 & 0.84574   \\
$r = 2, d=5$   && 0.29993 & 0.33348 & 0.39512 & 0.32317 & 0.89940 & 1.18484 & 0.96909   \\
$r = 2, d=6$   && 0.30055 & 0.31620 & 0.39959 & 0.32357 & 0.95048 & 1.26372 & 1.02329    \\  \\

  &\multicolumn{8}{c}{Panel (c): $\beta_1 \in (-1,0), \beta_2 \in (0,1)$} \\

$r = 0.8, d=4$   && 0.30001 & 0.41772 & 0.26456 & 0.33607 & 0.71820 & 0.63335 & 0.80454   \\
$r = 0.8, d=5$   && 0.29908 & 0.37058 & 0.26511 & 0.33689 & 0.80706 & 0.71538 & 0.90910   \\
$r = 0.8, d=6$   && 0.30038 & 0.35256 & 0.26684 & 0.33917 & 0.85199 & 0.75686 & 0.96201    \\  \\

$r = 2, d=4$   && 0.30012 & 0.41655 & 0.27130 & 0.33070 & 0.72051 & 0.65132 & 0.79392   \\
$r = 2, d=5$   && 0.30027 & 0.37022 & 0.27162 & 0.33141 & 0.81107 & 0.73367 & 0.89518    \\
$r = 2, d=6$   && 0.30058 & 0.35262 & 0.27223 & 0.33241 & 0.85242 & 0.77202 & 0.94269    \\  \\

  &\multicolumn{8}{c}{Panel (d): $\beta_1 \in (0,1), \beta_2 \in (-1,0)$} \\

$r = 0.8, d=4$   && 0.29979 & 0.34089 & 0.48870 & 0.32159 & 0.87942 & 1.43359 & 0.94338    \\
$r = 0.8, d=5$   && 0.29994 & 0.30282 & 0.52276 & 0.32426 & 0.99050 & 1.72634 & 1.07081   \\
$r = 0.8, d=6$   && 0.30005 & 0.29188 & 0.53985 & 0.32555 & 1.02798 & 1.84959 & 1.11535    \\  \\

$r = 2, d=4$   && 0.29978 & 0.34135 & 0.52265 & 0.31501 & 0.87822 & 1.53112 & 0.92284   \\
$r = 2, d=5$   && 0.30037 & 0.30242 & 0.56744 & 0.31838 & 0.99322 & 1.87633 & 1.05277  \\
$r = 2, d=6$   && 0.29971 & 0.29286 & 0.59408 & 0.32005 & 1.02337 & 2.02852 & 1.09284    \\  \\

   \hline
\end{tabular}
}

\end{table}

\section{Conclusions}
\label{Conclusions}

This paper addresses the problem of one-step-ahead forecast of an AR(1) process with a single structural break at an  unknown time and of unknown sign and magnitude within a very short learning sequence.
We analysed, via simulation experiments,
the forecast performance
of a smoothed predicting kernel algorithm
relative to that of a linear predictor that utilize the AR(1) model parameter estimated from the learning sequence without taking into account the presence of structural break. \par

It appears that
the shorter the learnings sequence,
the better
the forecast performance of
the smoothed predicting kernel
relative to
the linear predictor.
Regardless whether
the innovation terms in the learning sequences are constructed from
IID random Gaussian variables,
IID random Gamma variables,
IID scaled pseudo-uniform variables,
or
first-order auto-correlated Gaussian process,
the forecast performance of
the smoothed predicting kernel
is better than that of
the linear predictor
if the AR(1) model parameter switches from a negative value to a positive value in the learning sequence, i.e.,
$\beta_1 \in (-1,0), \beta_2 \in (0,1)$.
However,
it is not so
for the other regime switching scenarios
considered in the simulation experiments, i.e.,
$\beta_1, \beta_2 \in (0,1)$,
$\beta_1, \beta_2 \in (-1,1)$, and
$\beta_1 \in (0,1), \beta_2 \in (0,1)$. \par

It could be interesting to
explore the forecast performance
of the smoothed linear predictor
in the context of
random-coefficient AR(1) process
\citep[see, among others,][]{Leipus:Paulauskas:Surgailis:2006}
where
the AR(1) model parameter between any two sequential
observations are independent and identically distributed random variables from the uniform distribution $U[0,1]$.
Additionally, we may
explore the implementing
the smoothed predicting linear predictor
in the context of adaptive linear filtering
to perform successive, on-line
one-step ahead
forecast.
We leave this for future work. \par


%

\end{document}